\documentclass[11pt]{article}
\usepackage{graphicx}
\usepackage{amsmath}
\usepackage{hhline}
\usepackage{amssymb}
\usepackage{times}
\usepackage{cite}
\usepackage{array}
\usepackage{multirow}

\usepackage{color}
\definecolor{rltred}{rgb}{0.75,0,0}
\definecolor{rltgreen}{rgb}{0,0.5,0}
\definecolor{rltblue}{rgb}{0,0,0.75}

\newlength{\dinwidth}
\newlength{\dinmargin}
\begin{document}
\begin{titlepage}

\noindent
Date:        25 October 2016       \\
                
\vspace{2cm}

\begin{center}

\begin{Large}
{\bf MuSIC : delivering the world's most intense muon beam}
\end{Large}

\vspace{2cm}

S. Cook$^1$, R. D'Arcy$^1$, A. Edmonds$^1$, M. Fukuda$^2$, K. Hatanaka$^2$, Y. Hino$^3$, \\
Y. Kuno$^3$, M. Lancaster$^1$, Y. Mori$^4$, T. Ogitsu$^5$, H. Sakamoto$^3$, A. Sato$^3$, \\
N.H. Tran$^3$, N.M. Truong$^3$, M. Wing$^1$, A. Yamamoto$^5$ and M. Yoshida$^5$

\vspace{2cm}
$^1$Department of Physics and Astronomy, UCL, Gower Street, London, WC1E 6BT, UK \\
$^2$Research Center for Nuclear Physics (RCNP), Osaka University, Osaka, Japan \\
$^3$Department of Physics, Graduate School of Science, Osaka University, Osaka, Japan \\
$^4$Kyoto University Reactor Research Institute (KURRI), Kyoto, Japan\\
$^5$High Energy Accelerator Research Organization (KEK), Tsukuba, Japan\\

\end{center}

\vspace{2cm}

\begin{abstract}
A new muon beamline, muon science innovative channel (MuSIC), was set up at the Research Center
for Nuclear Physics (RCNP), Osaka University, 
in Osaka, Japan, using the 392\,MeV proton beam impinging on a target.  
The production of an intense muon beam relies on 
the efficient capture of pions, which subsequently decay to muons, using a novel superconducting 
solenoid magnet system.  After the pion-capture solenoid the first $36^\circ$ of the curved 
muon transport 
line was commissioned and the muon flux was measured.  In order to detect muons, 
a target of either copper or magnesium 
was placed to stop muons at the end of the muon beamline. Two stations of 
plastic scintillators located upstream and downstream from the muon target 
were used to reconstruct the decay spectrum of muons.  In a complementary method 
to detect negatively-charged muons,  
the X-ray spectrum 
yielded by muonic atoms in the target 
were measured in a germanium detector.  Measurements, at a 
proton beam current of 6\,pA, yielded $(10.4 \pm 2.7) \times 10^5$ muons per Watt of proton beam power 
($\mu^+$ and $\mu^-$), 
far in excess of other facilities.  At full beam power (400\,W), this implies a rate of muons 
of $(4.2 \pm 1.1) \times 10^8$\,muons\,s$^{-1}$, amongst the highest in the world.  The number of $\mu^-$ 
measured was about a factor of 10 lower, again by far the most efficient muon beam produced.  
The set up is a prototype for 
future experiments requiring a high-intensity muon beam, such as a muon collider or neutrino 
factory, or the search for rare muon decays which would be a signature for phenomena beyond 
the Standard Model of particle physics.  Such a muon beam can also be used in other branches of 
physics, nuclear and condensed matter, as well as other areas of scientific research.
\end{abstract}
\end{titlepage}

\section{Introduction}

High intensity muon beams have applications in many areas of science, spanning high energy 
particle physics to condensed matter physics and even areas of chemistry 
and biology.  Many results are limited by statistics and, depending on the experiment, up to and above 
$10^{18}$\,muons per year are required, whereas only $10^{15}$\,muons per year are available now.

In particle physics, intense muon beams are needed for the following experiments and areas of 
investigation. Rare muon decays such as charged lepton flavour violation (CLFV) have attracted much attention theoretically and experimentally~\cite{kunokada,lfv-review}. 
As the Standard Model (SM) expectation for such processes is so small ($\sim O(10^{-54}$)), higher intensity muon beams could lead to the unequivocal discovery of physics beyond the SM.  There are several current and planned experiments searching for CLFV with muons. They are
for example $\mu \to e \gamma$~\cite{mu-e-gamma}, $\mu - e$ conversion in a muonic atom~\cite{comet,mu2e} and $\mu \to eee$~\cite{mu3e}.  In particular, planned experiments of COMET~\cite{comet} in Japan and Mu2e~\cite{mu2e} in the US, which will search for $\mu - e$ conversion with anticipated improvement of physics sensitivity of $10^{4}$, need high intensity muon beams of $10^{18}$\,muons per year. The properties of the muon such as its mean lifetime, which gives a direct determination 
of the Fermi constant, or anomalous magnetic moment have both been measured to a precision of about 
one part per million~\cite{mu-lifetime,g-2,pdg}.  Given the approximate $3\,\sigma$ difference between 
theory and data in the measurement~\cite{g-2} of the anomalous magnetic moment, new experiments to measure with a factor of four better precision are currently under construction~\cite{new-g-2,jparc-g-2}.   Highly intense muon beams of $10^{21}$\,muons per year are needed for a muon collider, a machine that can investigate the energy frontier, i.e.\ the TeV scale.  A muon collider has a number of advantages such as compactness and lower synchrotron radiation compared to an $e^+e^-$ collider but also has a number of technical challenges~\cite{muon-collider}.

The use of muon beams in condensed matter physics, in particular as probes of the magnetic properties of 
materials, is given in detail elsewhere~\cite{cond-matter1,cond-matter2}.  Muons from the decay of charged 
pions at rest are naturally 100\% polarised and are subsequently stopped 
by the material under investigation. Internal magnetic fields inside the material can be studied by precession of the muon spin, which can be detected from the time-dependent angular distribution of the emitted electrons, and 
so does not involve scattering 
as neutron or X-ray material probes do.  Also, X-ray emission spectra from negative muons captured in matter 
provide a non-destructive method of determining the elemental content of a given sample, e.g.\ the 
characterisation of an archaeological artefact~\cite{x-ray-elements1,x-ray-elements2}.

Muon beams are usually produced via the decay of a large number of charged pions, produced 
by colliding a proton with a fixed target.  The challenges of producing high intensity muon beams are\,: the 
need for a high power proton beam; the efficient capture of pions produced at the target; and, given the large 
number of particles produced with a wide range of kinematic properties, effective methods to achieve a pure 
muon beam.  
Often, muon facilities and neutron facilities are combined. The pion production target for a muon beam requires small beam loss as a neutron production target is located downstream; in such cases, a 
short target and no magnetic field surrounding the target are employed.  At MuSIC, which is a dedicated muon source, a relatively long target is used 
and coupled with a system of superconducting solenoid magnets~\cite{music-magnets}, a high muon production efficiency is achieved.  The scheme to capture pions using solenoid magnets was first discussed by the MELC experiment~\cite{melc1,melc2} 
and is also proposed for muon conversion experiments~\cite{comet,mu2e}, muon colliders~\cite{muon-collider} and 
neutrino factories~\cite{nu-fact}.
Additionally, unlike particles usually used in particle accelerators, muons have a finite lifetime of 
2.2\,$\mu$s and so methods are needed to store them before decay.   The most intense muon beam produced 
is the $\mu$E4 beamline at the Paul Scherrer Institut in Switzerland which is capable of producing 
$J_\mu = 4 \times 10^8$\,muons\,s$^{-1}$ with momenta of about 28\,MeV~\cite{psi-beam}.  Given an 
initial proton beam power of $W_p = 1.2$\,MW, this equates to $J_\mu/W_p = 3.5 \times 10^{2}$\,muons\,s$^{-1 }$\,W$^{-1}$.

In this article, a new high intensity muon beamline, muon science innovative channel (MuSIC), at the Research 
Center for Nuclear Physics (RCNP), Osaka University, in Osaka, Japan is presented.  The article reports on the commissioning of the 
beamline and a measure of the intensity of muons produced; the facility can be used for much of the variety of 
science discussed above.  

\section{High intensity muon source}

High intensity muon sources require the collection of many pions which will produce muons in their decays.
To collect as many pions (and cloud muons) as possible, the pions are captured using a 
high-strength solenoidal magnetic field giving a large solid angle acceptance. The pion 
capture system consists of the pion production target, high-field solenoid magnets for 
pion capture, and a radiation shield.  In the MuSIC case, pions emitted into the backward 
hemisphere can be captured within a transverse momentum threshold,
$p_\mathrm{T}^\mathrm{max}$.
This $p_\mathrm{T}^\mathrm{max}$ is given by the magnetic field
strength, $B$, and the radius of the inner bore of solenoid magnet,
$R$, as
\begin{equation}
p_\mathrm{T}^\mathrm{max} (\textrm{GeV}/c) =
	0.3 \times B ({\rm T}) \times \frac{R({\rm m})}{2}.
\label{eq:max_pt-main}
\end{equation}
The target is located at the position of the maximum magnetic field to maximise the solid angle
for pions.  It is known that the higher the pion capture magnetic field,
the better the muon yield at the exit of the muon beamline. In
addition, the beam emittance is better if a higher magnetic field is
used for pion capture.  Therefore a higher magnetic field is preferable.  

The pions captured at the pion capture system have a broad directional distribution. In order to
increase the acceptance of the muon beamline it is desirable to make them more parallel to the beam
axis by changing the magnetic field adiabatically.  
From the angular momentum conservation, under a solenoidal magnetic field, the
product of the radius of curvature, $R$, and the transverse momentum, $p_{\rm T}$, is an invariant:
\begin{equation}
p_{\rm T}\times R \propto \frac{p_{\rm T}^2}{B} = \mathrm{constant,}
\end{equation}
where $B$ is the magnitude of the magnetic field.  Therefore, if the magnetic field decreases
gradually, $p_{\rm T}$ also decreases, yielding a more parallel beam.  This is the principle of the
adiabatic transition.  Quantitatively, when the magnetic field is reduced by a factor of 1.75 (as for the MuSIC case where a magnetic field of 3.5 T is changed into 2 T), 
$p_{\rm T}$ decreases by a factor of $\sqrt{1.75}$. On the other hand, since
\begin{equation}
 p_{\rm T}\times R \propto B\times R^2 =
 {\rm constant'},
\end{equation}
the radius of curvature increases by a factor of $\sqrt{1.75}$.
Therefore, the inner radius of the magnet in the muon transport section has
to be $\sqrt{1.75}$ times that of the pion capture solenoid (or more precisely the
inner radius of the radiation shielding of the pion capture solenoid).  At the
cost of an increased beam size, the pion beam can be made more parallel.

The selection of an electric charge and momenta of beam particles can be performed by using curved
(toroidal) solenoids, which makes the beam dispersive. A charged particle in a solenoidal field will
follow a helical trajectory. In a curved solenoid, the central axis of this trajectory drifts in the
direction perpendicular to the plane of curvature.  The magnitude of this drift, $D$, is given by
\begin{eqnarray}
D &=& \frac{1}{q  B} \left( \frac{s}{R} \right)
\frac{p_{\rm L}^2 + \frac{1}{2}p_{\rm T}^2}{p_{\rm L}}, \\
& = &
\frac{1}{q B} \left( \frac{s}{R} \right)
\frac{p}{2}\left( \cos\theta + \frac{1}{\cos\theta}\right) \,,
\end{eqnarray}
where $q$ is the electric charge of the particle (with its sign), $B$ is the magnetic field at the
axis, and $s$ and $R$ are the path length and the radius of curvature of the curved solenoid,
respectively.  Here, $s/R$ ($=\theta_{\rm bend}$) is the total bending angle of the solenoid, hence $D$
is proportional to $\theta_{\rm bend}$.  The quantities $p_{\rm L}$ and $p_{\rm T}$ are longitudinal and transverse momenta where 
$\theta$ is the pitch angle of the helical trajectory.  Because of the dependence on $q$, charged
particles with opposite signs move in opposite directions.  This can be used for charge and momentum
selection if a suitable collimator is placed after the curved solenoid.

To keep the centre of the helical trajectories of muons with a reference momentum $p_0$ in the
bending plane, a correction dipole (CD) field parallel to the drift direction can be applied. If a
correction dipole field, $B_{\mbox{\scriptsize CD}}$, given by
\begin{equation}
B_{\mbox{\scriptsize CD}} = 
\frac{1}{q R}\frac{p_0}{2}
\left(\cos\theta_0 + \frac{1}{\cos\theta_0}\right) \, ,
\end{equation}
is applied, the trajectories of particles of charge $q$ with momentum $p_0$ and pitch angle
$\theta_0$ will be corrected to be on-axis.

\section{Muon production at RCNP}

The proton cyclotron accelerator at RCNP, Osaka University, 
Japan, provides a continuous proton beam~\cite{rcnp}.  
Protons are produced in an ion source and accelerated in two stages, initially up to about 
65\,MeV, and finally up to about 400\,MeV.  A maximum beam current of 1\,$\mu$A, corresponding to a 
beam power of 400\,W, can be transported to the MuSIC experimental facility.  The results 
presented here are based on a beam energy of 392\,MeV and a proton beam 
current, $I_p$, in the range 
from 6\,pA to 1\,nA measured during data taking using monitors at the end of the beamline.  

\begin{figure}[htp]
\includegraphics[width=\textwidth]{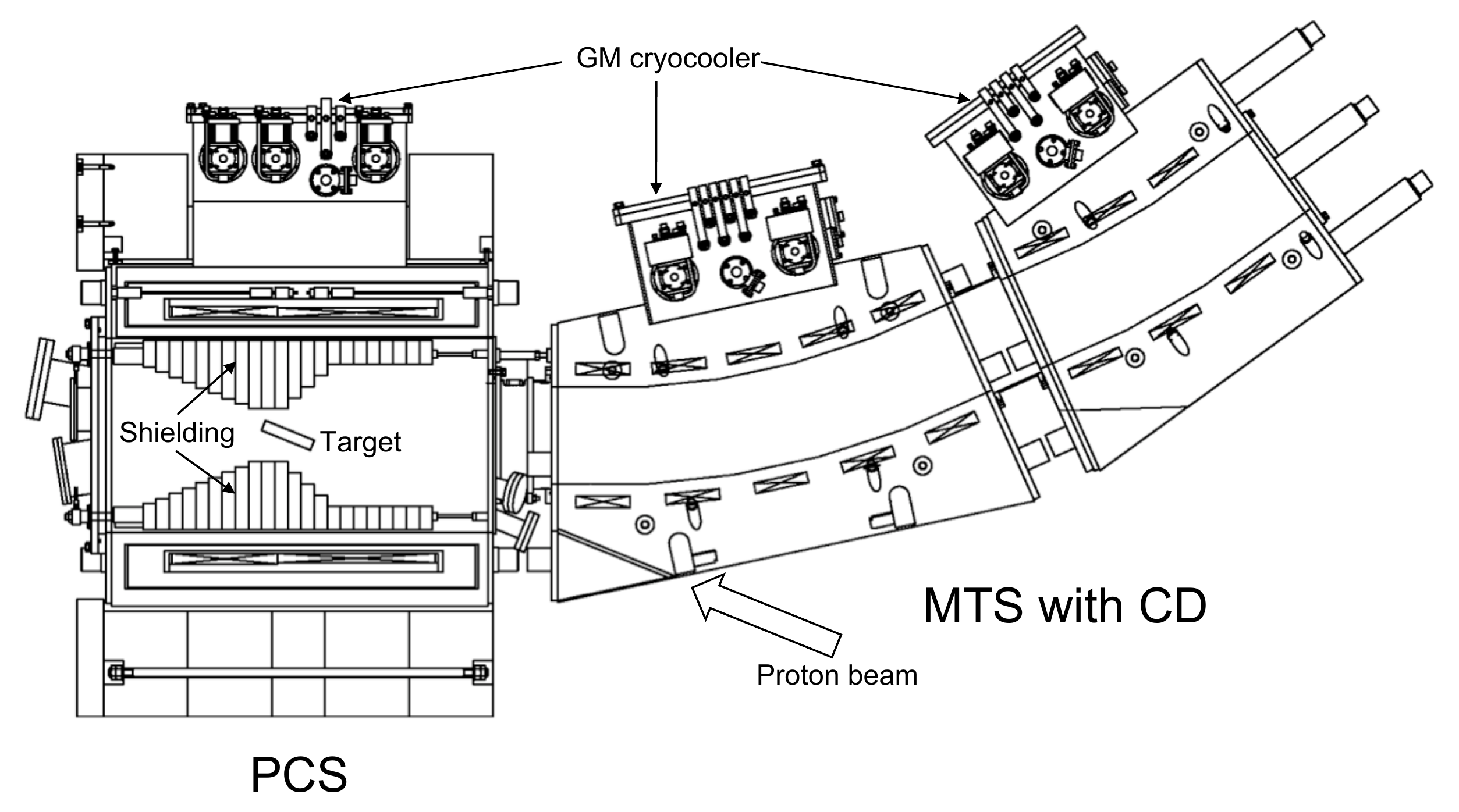}
\caption{Layout of the MuSIC beamline showing the target surrounded by shielding and the pion capture solenoid 
(PCS).  To the right are the two modules of five and three element coils, respectively, of the muon transport solenoid 
(MTS), covering an arc of $36^\circ$.  Each coil in the MTS includes a correction dipole (CD).   All superconducting 
coils are cooled with conduction cooling using Gifford--McMahon (GM) cryocoolers.  The direction of the 
incoming proton beam is also indicated.}
\label{fig:magnets}
\end{figure}

The proton beam impacts a graphite cylindrical target of length 20\,cm and radius 2\,cm, at an angle 22$^\circ$ 
horizontally from the surrounding pion capture solenoid (PCS) axis, see Figure~\ref{fig:magnets}, so that the proton 
beam trajectory 
and the target axis are aligned.  
Fluorescent plates are attached to both ends of the target circular surface so 
that, by looking at the fluorescent light, the beam can be centred on the target face.  The target is supported by a 
support shaft of 5\,m in length and can be removed or inserted easily.  The target is surrounded by stainless 
steel shielding of up to 27\,cm thick, tapering on either side of the target.  The taper is more rapid in the 
backwards direction, see Figure~\ref{fig:magnets}, in order to capture the maximum number of pions and muons.

To reduce backgrounds from in particular neutrons and protons which will primarily be emitted in the forward (direction 
of initial proton beam) direction, backward-going pions are collected.  Backward-going pions are captured by the PCS 
with a peak magnetic field of 3.5\,T and focused towards the muon transport solenoid (MTS) via the graded magnetic 
field.  The design parameters of the PCS are given in Table~\ref{tab:magnets}~\cite{music-magnets}.  

The copper-stabilised NbTi superconducting coils are cooled using three GM cryocoolers which have a total cooling 
power of 4\,W at the operating temperature of 4\,K.  Nuclear heating, mostly due to neutrons, was estimated using the 
MARS simulation~\cite{mars1,mars2,mars3,mars4,mars5}.  The coil density was assumed to be 9\,g$/$cm$^3$ and the 
thickness of the stainless steel support structure of the coil was 10\,mm.  The total energy deposited at the solenoid 
coils including the support structure was about 0.6\,W for a proton beam of 400\,MeV and 1\,$\mu$A.  Given a static 
heat load of 1\,W, the combined value is below the cryocoolers total cooling power.

\begin{table}[htp]
\caption{Design parameters of the pion capture solenoid (PCS), muon transport solenoid (MTS) and correction dipole 
(CD) magnets.  The coil thickness for the PCS is 40\,mm for upstream of 400\,mm to increase the magnetic field. RRR is 
a residual-resistance ratio of superconductors used.} 
\label{tab:magnets}
	\begin{center}
	\begin{tabular}{|c||c|c|c|} \hline 
	& PCS & MTS & CD \\ \hline \hline
	Conductor & Cu stabilised NbTi & Cu stabilised NbTi & Cu stabilised NbTi \\ \hline
	Conductor diameter (mm) & 1.2 & 1.2 & 1.2 \\ \hline 
	Cu / NbTi & 4 & 4 & 4 \\ \hline
	RRR (R293K / R10K) & $>$ 240 & $> $ 150 & $>$ 150 \\ \hline 
	Coil diameter (mm) & 900 & 480 & 460\\ \hline
	Coil length (mm) & 1\,000 & 200 & 200 \\ \hline
 	Coil thickness (mm) & 35 & 30 &  \\ \hline
	Number of turns & 30\,000 & 4\,000 & 528 \\ \hline
	Operation current (A) & 145 & 145 & 115 (Bipolar) \\ \hline
	Field (T) & 3.5 & 2 & 0.04 \\ \hline
	Inductance (H) & 400 & 124 & 0.04 \\ \hline
	Stored energy (MJ) & 5 & 1.4 & \\ \hline
	Quench back heater & 1.2\,mm Cu wire & 1.3\,mm Cu wire & \\ \hline
	 \end{tabular}
	\end{center}
\end{table}

The curved MTS is employed to preferentially select muons in the momentum region 
($\sim 20-60$\,MeV) low enough to be subsequently stopped in a target further downstream.  In order to have a 
high transport efficiency for such muons, the solenoid has a large bore of 36\,cm with an on-axis magnetic field 
of 2\,T.  A correction dipole (CD) magnetic field from $-0.04$\,T to $+0.04$\,T is applied in order to keep the trajectory of the low 
energy muons roughly centred and to filter out other particles. 
The CD magnetic field can be also used to select electric charges of muons in the beam coming to downstream. However, the covered arc of 36$^{\circ}$ is not large enough to obtain good separation of charges, and therefore, contamination of oppositely charged particles in the beam, could not be fully eliminated, as mentioned later.  
The design parameters of the MTS and CD are given in Table~\ref{tab:magnets}.

\section{Muon detection}
\label{sec:detector}

The particles exiting the beam-pipe were a mixture of electrons and positrons, pions and muons of both charges and 
protons and neutrons, because of the short arc of 36$^{\circ}$ coverage of the MTS.  The challenge of this experiment is to measure the muon flux above such 
high backgrounds.  Two methods were employed, the first of which measures electrons (and positrons) from muon decays
and the second 
measures negatively charged muons.  The first relies on measuring the decay time spectrum of the muon to 
identify and count their number by measuring a coincidence of a muon in one detector followed by an electron in a later 
detector.  The second measures the muonic X-ray energy spectrum emitted by the negatively-charged muons captured by a nucleus.  The experimental set-ups are given in the following.

\subsection{Muon decay spectrum}

A system to measure the muon flux was made by detecting electrons (and positrons) from muon decays.
The correction magnetic field $B_{\rm CD}=-$0.04\,T was set to primarily select positively charged muons.
According to simulations (see Section~\ref{sec:simulation}), electrons made up about 74\% of the beam with positrons the next 
most prevalent, making up about 10\% of total population.  The fraction of positive and negative muons was estimated 
to be about 5\% and 0.5\% respectively

\begin{figure}[ht]
\includegraphics[width=\textwidth]{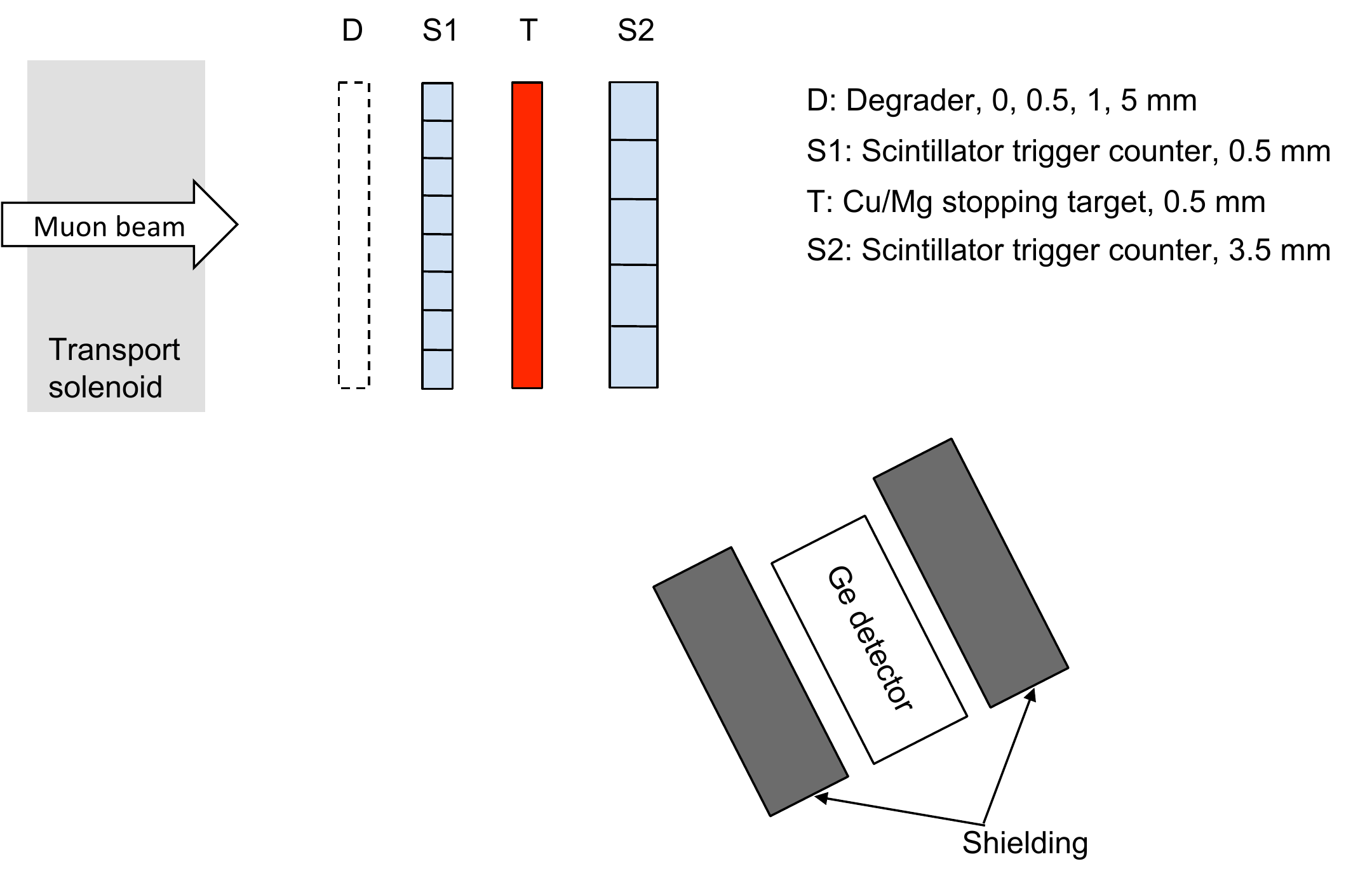}
\caption{Schematic of the independent detection systems for measuring the muon flux.  After 36$^\circ$ arc of the
muon beamline, 
the beam, including muons, exits the transport solenoid and optionally passes through an aluminium degrader, D.  The 
beam then passes through a plastic scintillator trigger counter, S1, and a target, T, where muons are stopped.  X-rays 
emitted by a captured muon dropping down orbital energy levels are detected in a germanium detector which is also 
shielded.  Electrons arising from the decay of a muon are registered in another plastic scintillator, S2.}
\label{fig:detector}
\end{figure}

An aluminium degrader, with height and width both of 400\,mm, was optionally placed immediately after the beam exited 
the beam-pipe and magnet system.  Thicknesses of 0.5, 1 and 5\,mm were used in order to select different ranges of muon momentum.  From simulations, the 
mean momentum for muons at the end of the beam-pipe and subsequently stopped were e.g.\ with no degrader and a 
5\,mm-thick degrader, $45.2 \pm 0.2$\,MeV and 
$66.1 \pm 0.2$\,MeV, respectively.  After the degrader, a copper or magnesium stopping target was sandwiched between 
two sets of plastic scintillator counters.  The first scintillator, S1, consisted of eight channels, with height and width of 
30\,mm and 380\,mm and was 0.5\,mm thick, in order to disturb the beam minimally.   This counter was used to 
detect the passage of the initial muons.  The second scintillator, S2, consisted of five channels, with height and width of 
50\,mm and 380\,mm, and was of thickness 3.5\,mm; this was used to detect electrons from the decay of a muon.  
The scintillators were wrapped in reflective mylar foils and black plastic sheets to prevent light leakage.  Each 
scintillator strip had a wavelength-shifting fibre mounted on the back of it which was connected to a multi-pixel photon 
counter (MPPC) at each end.  The signals from each pair of MPPCs were combined and amplified before being passed 
to the data acquisition (DAQ) system for processing.  A schematic of the set-up is shown in Figure~\ref{fig:detector}.

In order to trigger on muon decays, a ``hit" recorded in S2 was required to occur 50\,ns to 20\,$\mu$s after the hit in S1.  
A hit in S1 was defined as a coincident signal in both MPPCs of a given scintillator strip where the signal value was at 
least 8\,photons, set by a discriminator.  A hit in S2 was given by a signal of at least 10\,photons, higher than in S1 as 
electrons deposit more energy in the thick scintillators than slow muons in the thin scintillators.  The time difference, 
$t$, between the signals in S1 and S2 was used for further analysis in order to reconstruct the muon decay time
spectrum.

\subsection{Muonic X-ray measurement} 

A complementary system to measure the negative muon flux was the use of a germanium detector to measure X-rays emitted by 
negatively charged muons as they drop down energy levels in a muonic atom 
after being captured by one of the nuclei in the target, T, shown in Figure~\ref{fig:detector}.
A stopping target was made of magnesium of 20\,mm thickness, with height and width of 80\,mm and 370\,mm.  The 
germanium detector is a planar type of Canberra GL0515R with a germanium crystal of an active diameter of 25.2\,mm and 
depth of 15\,mm. The cryostat window, made of beryllium, was 0.15\,mm thick.  To ensure low backgrounds from particles 
exiting the beam line and other secondary interactions, the detector was placed off axis by 25$^{\circ}$ from the beam 
axis and a distance of 500\,mm from the target. The CD magnetic field of $B_{\rm CD}=+0.04$\,T was chosen to 
select negatively charged muons. This setup is also shown schematically in Figure~\ref{fig:detector}.  

Since the detector was not far from the proton target, the neutron flux at the detector location was very high. To reduce 
neutrons, the germanium detector was shielded with paraffin, cadmium, and lead blocks from outer to inner sides. The 
paraffin cylindrical shielding of 100\,mm thickness was used to decelerate fast neutrons to thermal neutrons. The 
cadmium shielding of 2\,mm thickness was used to absorb thermal neutrons followed by $\gamma$-ray emission. The 
lead shielding of 50\,mm thickness was used to absorb $\gamma$-rays from cadmium and other $\gamma$-rays.  
Furthermore, it was found that characteristic X-rays from lead blocked the muonic $L_{\alpha}$ X-rays and the pionic 
X-rays from magnesium.  Additional shielding, which consisted of cylindrical tubes of tin, copper and aluminium (outer 
to inner) were placed between the germanium detector and the lead shielding. 

The signal output from the germanium detector was amplified and fed to a multi-channel analyser (MCA). The MCA was triggered by the muon stop logic signal formed by the beam hodoscopes.  The energy calibration of the germanium detector was made by using a $^{133}$Ba source.

\section{Simulation}
\label{sec:simulation}

The experimental set up was simulated in order to aid the design of the detectors, to determine efficiencies and 
acceptances, and to compare to the data.    Two codes were used to perform the simulation: 
{\sc G4Beamline}~\cite{g4bl} was used to simulate the hadron production and track particles through the beamline; and 
{\sc Geant4}~\cite{geant4} was used to simulate the detectors.

{\sc G4Beamline} was used to simulate the bulk of the experiment: the initial proton beam; the pion capture system, 
including the target, capture solenoid, shielding and return yoke; and the transport system, bending magnets and 
beampipe sections.  The position and momentum of all particles passing through the end of the beamline were recorded 
and used as input to the subsequent {\sc Geant4} simulation.

{\sc Geant4} is used to simulate the detector set-up shown in Figure~\ref{fig:detector}.  Using the output from {\sc G4Beamline}, 
the particles pass through materials constituting the various detector components, interacting with them according to  
formulae for energy loss, emission spectra, etc..  A full optical simulation of photon production in the scintillators and detection 
in the MPPCs was performed.  The QGSP\_BERT\_HP physics list in {\sc Geant4} was used to simulate hadronic interactions 
as this is designed to transport neutrons with energies as low as 20\,MeV.  The reconstruction and cuts described in 
Section~\ref{sec:detector} as well as the signal extraction described in Section~\ref{sec:analysis} were performed on the 
simulated data as for the real data.

\section{Analysis method}
\label{sec:analysis}

\subsection{Analysis of muon lifetime}

\begin{figure}[htp]
\includegraphics[width=\textwidth]{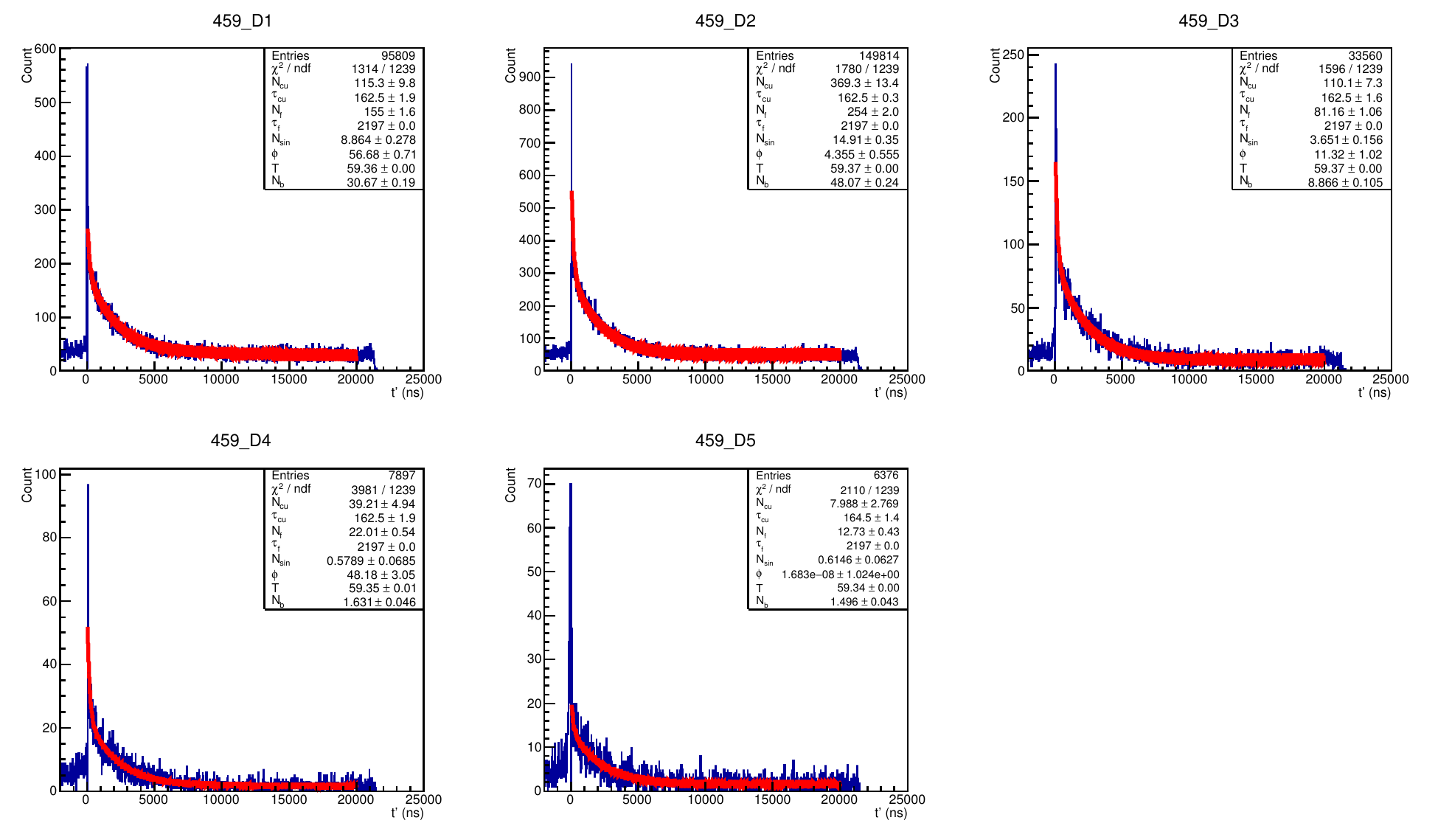}
\caption{Time distribution for an example data sample, with a degrader thickness of 5\,mm, for each of the 
five S2 channels.  The fit to the data with the function in Equation~(\ref{eq:fit}) and the parameters extracted for 
each distribution are also shown.}
\label{fig:time-fit}
\end{figure}

The distributions of $t$ are shown in Figure~\ref{fig:time-fit} for each of the five S2 channels for an example data 
sample (see~\cite{sam:thesis} for other samples).  Also shown is a fit, $N(t)$, to the data given by:

\begin{equation}
N(t) = N_f \exp \left( -\frac{t}{\tau_f} \right) + N_s \exp \left( -\frac{t}{\tau_s} \right)
      + N_{b1} \sin \left( 2\pi \frac{t - \phi}{T} \right) + N_{b2}\,,
\label{eq:fit}
\end{equation}
where $t$ is the time in ns.  The four components correspond to the free decay of positive muons, $f$, 
the decay of negative muons in the stopping target, $s$, a sinusoidal background, $b1$, and a 
flat background, $b2$.  The muon decays rates are parametrised with a scale factor, $N$, and a 
lifetime, $\tau$.  The sinusoidal background term, which comes from beam particles (mostly electrons) directly hitting
the counters, 
 has a period, $T$, and a phase relative to the trigger time, $\phi$.  
It is due to the minor bunching of the protons in acceleration with RF frequency of the RCNP cyclotron.
The flat background term is due to combinatorial background of 
beam particles which fake a signal.  Known values of the lifetimes, 
$\tau_f = 2\,196.9811 \pm 0.0022$\,ns~\cite{pdg} and, for a copper stopping target, the lifetime of a muonic atom of copper~\cite{cu-life}, $\tau_s = 163.5 \pm 1.0$\,ns, were fixed in the fit.  The value of $T$ was also fixed 
to $60 \pm 5$\,ns determined by fitting the background noise in a region where no signal is expected, i.e.\ 
high $t$.  The time distribution for an example data sample, with a degrader thickness of 5\,mm, is shown 
for each of the five S2 channels in Figure~\ref{fig:time-fit}.  The fit to the data with the function in Equation~(\ref{eq:fit} )
is also shown.  The fits to the data are reasonable with some distributions having values of $\chi^2$ per degree 
of freedom of about one 
as is the case for the first channel shown here.  Channels 4 and 5 give consistently poor 
values, worse than channels from 1 to 3, and so are excluded from further analysis.

The functions for the free decays of positive muons and decays of negative muons in 
a muonic atom of copper with parameters extracted in the fit to the data are then integrated 
in order to determine the number of muons in each sample.  As a cross check of the method, simulated distributions 
were fit with the function in Equation~(\ref{eq:fit}) and the number of muons extracted.  These values agreed with simply 
counting the number of real muons in the simulation.

The dead time of the DAQ system, 
due to it being busy, was calculated for each running configuration from the number 
of potential and good triggers.  The potential triggers were those with a signal in the 
upstream scintillator and no corresponding signal downstream within the 50\,ns veto window.  A good trigger was 
defined as a potential trigger without the system being busy.  The dead time varied from $38.43 \pm 0.03$\,\% when 
no degrader was used to $15.05 \pm 0.12$\,\% when a 5\,mm-thick degrader was used.  The decreasing dead time 
with increasing degrader thickness is to be expected due to the decrease in the overall rate of beam particles entering 
the detector system.  In order to calculate the muon flux, the data were corrected for the dead time.

In order to extract a measured value, independent of the detector set-up, the muon flux was corrected for the 
detector acceptance and the MPPC efficiency.  The 
detector acceptance was calculated with simulation by determining the 
fraction of muons which produced the signals described previously in the two scintillator detectors.  The value was 
$A = 79.6 \pm 1.2$\%.  The efficiency of the MPPCs was determined using cosmic rays by placing the scintillator 
and MPPC system between a set of large scintillator paddles with the scintillation photons detected by 
photomultiplier tubes.  By requiring a coincidence in the paddles above and below, the efficiency of an MPPC, 
accounting for the geometrical acceptance, was determined to be $\epsilon = 0.431 \pm 0.134$.  The large uncertainty comes 
from the lack of stability over time and the differences for different MPPCs.  Given the requirement in the 
measurement of signals in both MPPCs for a given scintillator strip, this gives an efficiency of 
$\epsilon^2 = 18.6 \pm 11.6$\%.  The efficiency of the MPPC system is the dominant systematic uncertainty.

Other uncertainties considered were related to the fitting procedure and 
extraction of the integrated number of the electrons and positrons from muon decays, $N_e$\,, in their time spectra.  The lower bound on the fit and the binning of the data was varied.  The uncertainty on the number of free positive 
muon decays was 
about 1.1\%, where for negative 
muons decaying in copper, it was about 56\%.  The large uncertainty for the results of decays in 
copper arise due to the sensitivity of the fit at small times where the decays in copper are concentrated.

The total number of muons stopped in the target, $N_{\mu}$, measured by the decay electrons and positrons is given by 
\begin{equation}
N_{\mu}= { N_e \over {A \cdot \epsilon^2} }\,,
\end{equation}
where $A$ and $\epsilon$ are the detector acceptance and the MPPC efficiency respectively. The data samples with the same degrader thickness give consistent results, demonstrating control of the analysis with time.  The muon rate also decreases with increasing degrader thickness, consistent with the expectation.  These two qualitative conclusions were observed for both free decays of positive 
muons and negative muons decaying in a muonic atom of copper. 

\subsection{Analysis of muonic X-ray measurement}

\begin{figure}[htb!]
	\begin{center}
\includegraphics[width=0.75\textwidth]{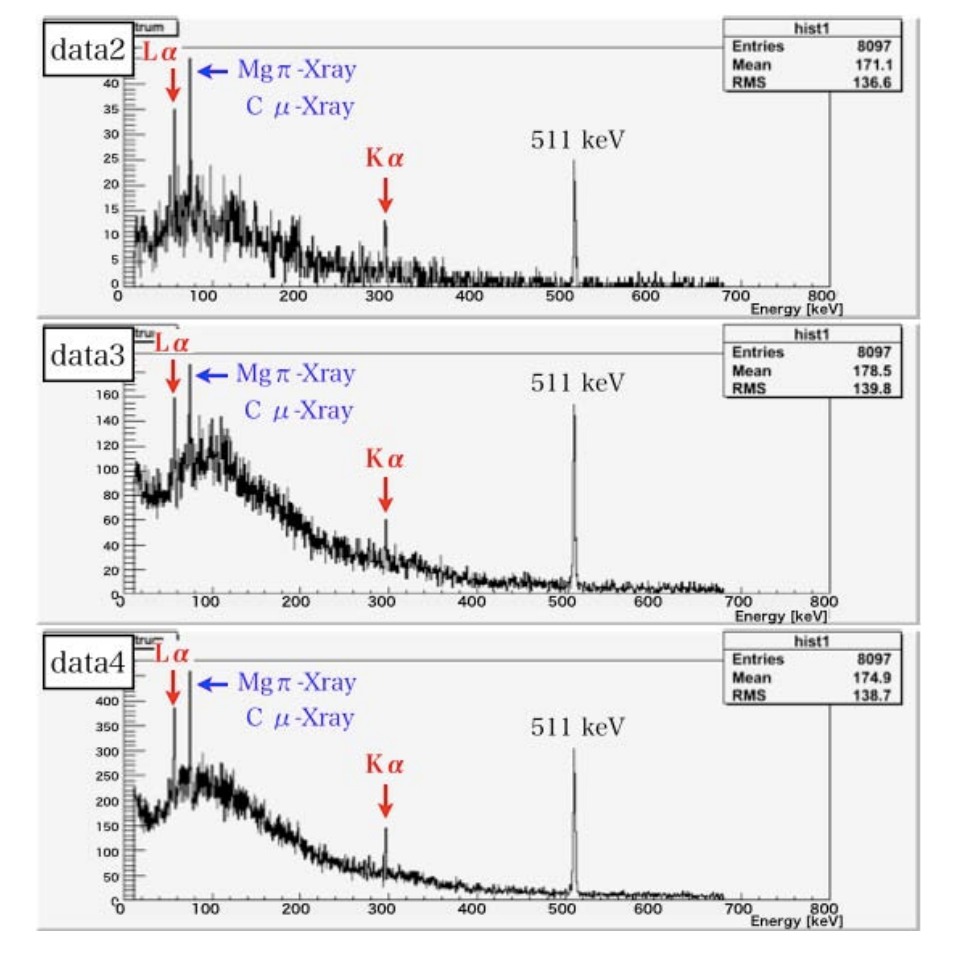}
	\end{center}
\caption{Muonic X-ray spectra for different data sets.}
\label{fg:x-ray-spectra}
\end{figure}

The muonic X-ray measurements on magnesium were done for three data sets \cite{hino2012}.  The list of the data sets is shown in Table~\ref{tb:datasets}, 
where different proton beam currents were used. The X-ray spectra for each data set are shown in Figure~\ref{fg:x-ray-spectra}, 
where muonic $K_{\alpha}$ and $L_{\alpha}$ X-rays were measured together with pionic X-rays.  The energy 
calibration was done using a $^{133}$Ba source.

\begin{table}[ht]
\caption{Data sets for muonic X-ray measurements with different proton beam currents.}
\label{tb:datasets}
\begin{center}
\begin{tabular}{|c|c|c|c|c|}\hline
Data & Measurement & Proton beam & Muonic $K_{\alpha}$ & Muonic $L_{\alpha}$ \cr 
name & time (s) & current (pA) & X-ray events & X-ray events \cr 
& & ($I_p$) & ($N_{2\rightarrow 1}$) & ($N_{3 \rightarrow 2}$) \cr\hline
data1 & $1.0\times 10^{4}$ & 59 & $23 \pm 6$ & $46 \pm13$ \cr \hline
data2 & $1.1\times 10^{4}$ & 134 & $72\pm12$ & $141\pm 22$ \cr \hline
data3 & $1.0\times 10^{4}$ & 435 & $212\pm21$ & $318\pm48$ \cr \hline
\end{tabular}
\end{center}
\end{table}

The muonic X-ray peaks were fit with a Gaussian and a constant background to obtain the total number of events of 
muonic X-rays. The energy dependence of the X-ray peak width was found from the measured data to be

\begin{equation}
\sigma = 0.0196 \sqrt{E ({\rm keV}) + 57}\,.
\end{equation}
The numbers of muonic X-ray events are summarised in Table~\ref{tb:datasets}.

The total number of negatively charged muons stopped in the target, $N_{\mu}$, was determined using the following

\begin{equation}
N_{\mu^-} = { N_{i \rightarrow j} \over P_{i \rightarrow j} \times (\Omega /4 \pi) \times \varepsilon }\,,
\label{eq:negativemuons}
\end{equation}
where $N_{i\rightarrow j}$ and $P_{i \rightarrow j}$ are the number and the emission probability of the muonic X-ray 
in transition from the muonic atomic state of $i$ to that of $j$; $P_{i \rightarrow j}$ is known~\cite{vogel80} and listed in Table~\ref{tb:x-ray-emission}.  The quantity $(\Omega / 4 \pi)$ is the solid angle of the germanium detector and 
$\varepsilon$ is the efficiency of the germanium detector. Their combination, $\Omega_{D} = (\Omega / 4\pi) \times \varepsilon$ can 
be estimated from {\sc Geant4} simulations taking into account the energy dependence of absorption by material and that of 
detection in germanium. The values $\Omega_{D}  = (1.28 \pm 0.05) \times 10^{-5}$ for $K_{\alpha}$ X-rays and 
$\Omega_{D} = (2.77 \pm 0.12) \times 10^{-5}$ for $L_{\alpha}$ X-rays were obtained.

\begin{table}[htb!]
\caption{Energy and emission probability for different muonic X-rays on magnesium.}
\label{tb:x-ray-emission}
\begin{center}
\begin{tabular}{|c|c|c|}\hline
Muonic X-ray & Energy& Probability\cr 
($i \rightarrow j$) & $E_{i \rightarrow j}$ (keV)~\cite{suzuki67} & $P_{i\rightarrow j}$ (\%)~\cite{vogel80} \cr \hline
$K_{\alpha}$ ($2 \rightarrow 1$) & 296.4 & $(79.6 \pm 0.7)$ \cr \hline
$L_{\alpha}$ ($3 \rightarrow 2$) & 56.6 & $(62.5 \pm 0.7)$ \cr\hline
\end{tabular}
\end{center}
\end{table}

The simulation was validated by the measurements with the standard calibration source of $^{133}$Ba with its known 
absolute strength. The source was placed 200\,mm away from the germanium detector. The comparison between the 
simulation and the measurement gave uncertainties of 2\,\% and 6\,\% for the reconstruction of the $K_{\alpha}$ and $L_{\alpha}$ 
X-rays.

\section{Results and measurement of muon beam intensity}

\subsection{Results from muon lifetime analysis}

\begin{figure}[htb!]
	\begin{center}
\includegraphics[width=0.7\textwidth]{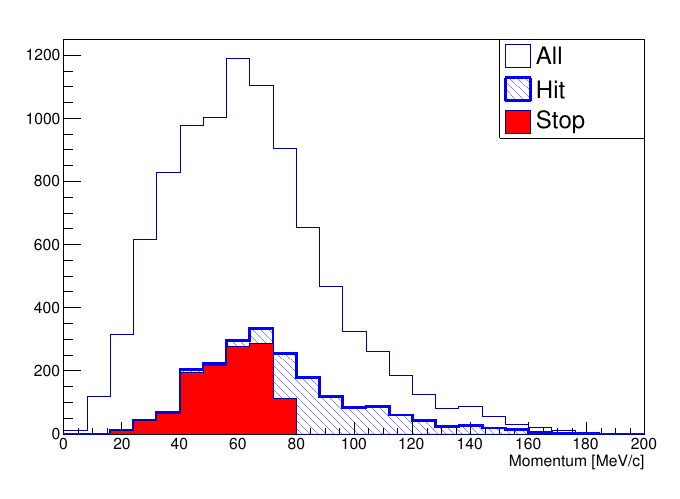}
	\end{center}
\caption{Momentum distributions for all muons in a beam (white), the muons hitting the trigger counters (blue) and those stopped in the target (red).}
\label{fg:momentum_stop}
\end{figure}

The final number needed in order to extract the total muon flux is the number of muons which pass through the 
degrader, subsequently stopped and are measured in the detector system.  
The simulation started from the 
pion production was used to get the muon stopping distribution in the stopping target. Figure~\ref{fg:momentum_stop} 
shows the momentum distribution of muons stopped in the target from the simulation.
From simulation, the probability of the muons stopping was determined to 
be at most $F_{s}=8$\%, depending on the degrader thickness and includes both positive and negative muons.  

Given an expected beam design current of 1\,$\mu$A, the final muon flux in muons per second is given by

\begin{equation}
J_{\mu}  = { N_{\mu} \over {I_{p} \cdot F_{s} \cdot D \cdot L} }\,,
\end{equation}
where $I_p$ is the proton beam current in $\mu$A, $D$ is the duration of the run and $L$ is the live time, i.e.\ the dead time subtracted from 100\%.  

The data samples taken with different degrader thickness are consistent with each other and so averaging over 
these values gives a final flux of $J_{\mu} = (4.2 \pm 1.1) \times 10^8$\,muons\,s$^{-1}$ which corresponds to an efficiency for 
full beam power (400\,W) of $J_{\mu}/W_p = (10.4 \pm 2.7) \times 10^5$\,muons\,s$^{-1 }$\,W$^{-1}$.   This value is far in excess of any 
other muon source, i.e.\ is the by far the most efficient muon beam produced. 

\subsection{Results from X-ray measurements}
 
From the three data sets with different proton beam currents, the muon beam intensity of negative muons, $J_{\mu^{-}}$\,, was obtained to be

\begin{equation}
J_{\mu^{-}}  = { N_{\mu^{-}} \over {I_{p} \cdot F_{s} \cdot D \cdot L} }\,,
\end{equation}
where $I_{p}$ is the proton beam current in $\mu$A and $D$ is the duration of the run and $L$ is the live time. $N_{\mu^{-}}$ is the total number of muons stopped in the target, given in Equation~(\ref{eq:negativemuons}). The measurements of $K_{\alpha}$ and $L_{\alpha}$ X-rays were 
combined. For a maximum proton beam current of 1 $\mu$A at RCNP, a muon yield of 
$J_{\mu^{-}} = (3.6 \pm 0.4) \times 10^{7}$\,muons\,s$^{-1}$, which is the world's highest, was obtained. The muon yield 
per beam power is $J/W_p = (9.0 \pm 1.0) \times 10^4$\,muons\,s$^{-1 }$\,W$^{-1}$, an improvement of about 1\,000 over 
existing facilities. {\sc Geant} simulations with {\sc QGSP\_BERT} give about 
$J_{\mu^{-}}  = 6.5 \times 10^{7}$\,muons\,s$^{-1}$, which is larger than the measured yield.  This could be due to the 
hadron code in {\sc Geant} simulations.

\section{Summary}

The world's most efficient muon beam has been produced at the MuSIC facility in Osaka, Japan.  The muon yield per beam 
power was measured to be (for $\mu^+$ and $\mu^-$) $(10.4 \pm 2.7) \times 10^5$\,muons\,s$^{-1 }$\,W$^{-1}$ and 
(for $\mu^-$ only) $(9.0 \pm 1.0) \times 10^4$\,muons\,s$^{-1 }$\,W$^{-1}$, over a factor of 1\,000 higher than current 
facilities.  Given a maximum beam power of 400\,W, rates of about $4 \times 10^8$\,muons\,s$^{-1}$ are achievable.  The 
increase in efficiency arises principally due to the use of a novel superconducting solenoid magnet system to capture pions 
produced at a target.  This demonstration in increased efficiency will be utilised in future muon experiments in order to maximise 
the flux of muons and hence search for charged lepton flavour violation.

\section{Acknowledgements}
This work is supported in part by the Japan Society of the Promotion of Science (JSPS) KAKENHI Grant No.\ 25000004. The support of the Science and Technology Facilities Council, UK is acknowledged.  M.~Wing acknowledges the support of DESY 
and the Alexander von Humboldt Stiftung.

\end{document}